\documentclass[final]{aipproc}
\layoutstyle{8x11double}

\begin{document}

\title{Propagation of Ultra-high-energy Cosmic Rays in Galactic Magnetic Field}


\author{Hajime Takami}{
  address={Max Planck Institute for Physics, F$\ddot{o}$hringer Ring 6, 80805 Munich, Germany}
}

\begin{abstract}
The propagation trajectories of ultra-high-energy cosmic rays (UHECRs) are inevitably affected by Galactic magnetic field (GMF). Because of the inevitability, the importance of the studies of the propagation in GMF have increased to interpret the results of recent UHECR experiments. This article reviews the effects of GMF to the propagation and arrival directions of UHECRs and introduces recent studies to constrain UHECR sources. 
\end{abstract}

\maketitle

%%%%%%%%%%%%%%%%%%%%%%%%%%%%%%%%%%%%%%%%%%%%%%%%%%
%%%%%%%%%%%%%%%%%%%%%%%%%%%%%%%%%%%%%%%%%%%%%%%%%%
\section{Introduction}
%%%%%%%%%%%%%%%%%%%%%%%%%%%%%%%%%%%%%%%%%%%%%%%%%%
%%%%%%%%%%%%%%%%%%%%%%%%%%%%%%%%%%%%%%%%%%%%%%%%%%

The origin of ultra-high-energy cosmic rays (UHECRs) has been an open problem in astrophysics for about 50 years, despite both theoretical and experimental efforts. A main difficulty to identify UHECR sources originates from the charge of UHECRs, which leads to the deflection of their propagation trajectories in Galactic magnetic field (GMF) and intergalactic magnetic field (IGMF).  These fields make UHECRs arriving at the Earth lose directional information on their sources contrary to neutral particles, e.g., photons and neutrinos. However, the deflection is expected to be small (several degree) at the highest energies ($\sim 10^{20}$ eV) even at an upper limit value of averaged IGMF, $B_{\rm IGMF} {l_{c,{\rm IGMF}}}^{1/2} < (1{\rm nG}) (1{\rm Mpc})^{1/2}$ \cite{Kronberg1994RepProgPhys57p325}, if UHECRs are protons. Recent $\gamma$-ray observations imply $\sim 10^{-15}$ G in voids \cite{Ando2010ApJ722L39}, although this estimation is controversial \cite{Neronov2011AA526A90}, and give conservative lower limits of IGMF in voids $10^{17-19}$ G \cite{Dermer2010arXiv1011.6660,Taylor2011arXiv1101.0932}.  Although lower values of these constraints seems to be positive to identify UHECR sources directly, structured IGMF, i.e., magnetic field in clusters of galaxies and filamentary structure, can significantly contribute to the deflections, which includes large uncertainty in the modelling \cite{Sigl2003PRD68p043002,Dolag2005JCAP01p009,Takami2006ApJ639p803,Das2008ApJ682p29,Kotera2008PRD77p123003}.

The propagation of UHECRs in GMF has also been studied from early days \cite{Osborne1973JPhA6p421,Berezinsky1979ICRC2p86}. In 1990s, the importance of GMF was revisited \cite{Stanev:1996qj,MedinaTanco:1997rt}, which was stimulated by potential anisotropy in the arrival distribution of UHECRs \cite{Stanev:1995my}. In addition, since UHECRs detected at the Earth are inevitably affected by GMF, the effect of GMF is essential to consider what UHECR sources are. Moreover, after the identification of several UHECR sources, UHECRs can be used as a background source to constrain local GMF structures.

The importance of considering the effect of GMF has increased recently. In order to interpret the correlation between UHECRs and extragalactic astrophysical objects reported by the Pierre Auger Observatory (PAO) \cite{Abraham2007Sci318p938} and to constrain UHECR sources, the propagation of UHECRs in Galactic space should be taken into account because of its inevitability. The properties of the propagation highly depends on the composition of UHECRs, which is still controversial between the PAO \cite{Abraham2010PRL104p091101} and High-Resolution Fly's Eye (HiRes) \cite{Abbasi2010PRL104p161101} [and Telescope Array (TA) (see Dr. Tameda's talk in this workshop)], since the deflections of trajectories of UHECRs are sensitive to their charge (e.g., Ref. \cite{MedinaTanco:1997rt}). Also, in the case of transient sources, GMF produces significant time-delay of UHECRs compared to neutral particles emitted at the same time \cite{Murase2008ApJ690L14}. 

This article reviews the effects of GMF to the propagation and arrival directions of UHECRs and introduces recent studies to constrain UHECR sources. 

%%%%%%%%%%%%%%%%%%%%%%%%%%%%%%%%%%%%%%%%%%%%%%%%%%
%%%%%%%%%%%%%%%%%%%%%%%%%%%%%%%%%%%%%%%%%%%%%%%%%%
\section{GMF modeling}
%%%%%%%%%%%%%%%%%%%%%%%%%%%%%%%%%%%%%%%%%%%%%%%%%%
%%%%%%%%%%%%%%%%%%%%%%%%%%%%%%%%%%%%%%%%%%%%%%%%%%

Here, we give readers a minimal summary on GMF modelling. See Dr. Han's talk in this workshop for recent progress on GMF observations and more details of modelling.

GMF is modelled by two components; magnetic fields in the Galactic disk and in the Galactic halo. Note that the halo field usually means magnetic field far from the Galactic disk. The disk field has been modelled by a bisymmetric spiral structure (BS) \cite{Sofue:1983aa}, an axisymmetric spiral structure (AS) \cite{Stanev:1996qj}, and concentric rings with field reversals (e.g. Refs. \cite{2005ApJ...619..297V,Sun2008AA477p573}). Which model can reproduce observations globally better is still a controversial issue \cite{Sun2008AA477p573,Pshirkov2011arXiv1103.0814}, but the existence of (at least one) field reversals has been reported (e.g. \cite{Han:2006ci}). There is not evidence that the directions of the field above and below the Galactic plane is different.

\begin{figure}
\includegraphics[clip,width=\linewidth]{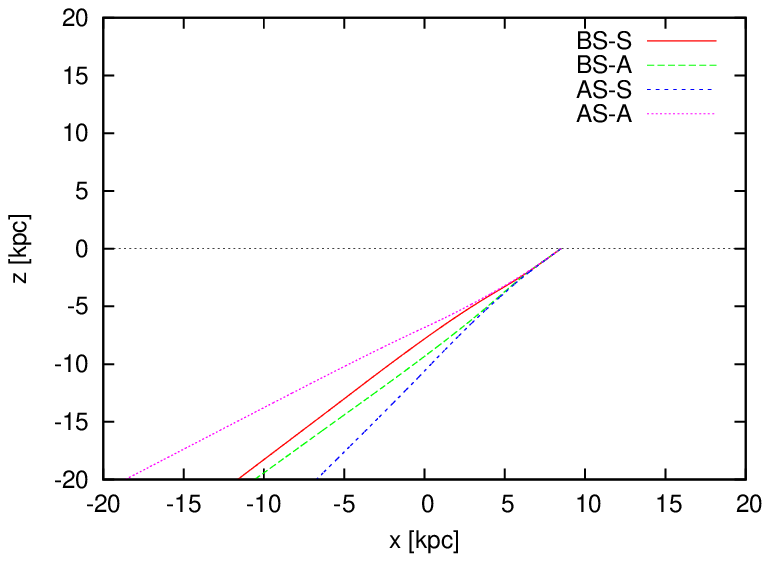} \\
\includegraphics[clip,width=\linewidth]{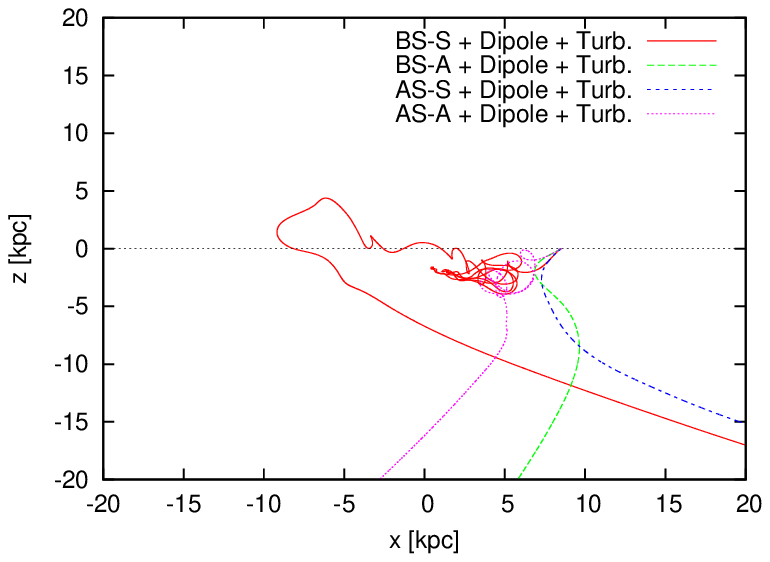}
\caption{Examples of the projected trajectories of UHECRs ({\it left}: protons, {\it right}: irons) with $10^{19.8}$ eV with the same arrival directions in different GMF models. This figure is originally from Ref.\cite{Takami2010ApJ724p1456}.}
\label{fig:trajectories}
\end{figure}

In the halo field, exponential decay models of the spiral disk field in the direction perpendicular to the Galactic plane have been used \cite{Stanev:1996qj,AlvarezMuniz:2001vf}. Faraday rotation measurements of extragalactic radio sources indicates the directions of magnetic fields parallel to the Galactic plane above and below the Galactic plane are opposite \cite{1997A&A...322...98H}. This anti-symmetric structure is theoretically supported by so-called A0 mode in dynamo theories. So, some models treat GMF in the disk and GMF in the halo separately \cite{Prouza:2003yf}. Recent observations have confirmed this treatment and have revealed the scale height of electron density in the disk higher than in previous understanding \cite{Sun2008AA477p573,Pshirkov2011arXiv1103.0814}.

A dipole field, as a GMF component parallel to the Galactic disk, is also often taken into account \cite{AlvarezMuniz:2001vf,Yoshiguchi:2003mc}. The dipole field is motivated by (i) a vertical field of 0.2-0.3 $\mu$G observed in the vicinity of the solar system \cite{1994A&A...288..759H}, (ii) strong filamentary magnetic field perpendicular to the Galactic plane in the vicinity of the Galactic center \cite{Han2007IAUS242p55} and (iii) a theoretical result of A0 dynamo, but there is no clear evidence of the dipole field. In extragalactic edge-on galaxies, X-shaped structures of magnetic field above and below the disk have been observed, which implies galactic wind \cite{Krause2007MmSAI78p314,Beck2009ApSS320p77}. The shape of the vertical component of GMF is still a controversial issue.

In addition to these coherent components, the turbulent component of GMF, whose strength is 0.5-2.0 times as strong as the coherent components, is implied \cite{Beck2000SSR99p243}. Only upper limits of the correlation length of the turbulent component ($\sim 100$ pc) have been estimated due to a limited angular resolution of radio observations \cite{Beck2000SSR99p243}. Although the turbulent component is not important for protons, it plays an significant role for heavy nuclei because the Larmor radius of heavy nuclei in the $\mu$G field approaches to the correlation length.

%%%%%%%%%%%%%%%%%%%%%%%%%%%%%%%%%%%%%%%%%%%%%%%%%%
%%%%%%%%%%%%%%%%%%%%%%%%%%%%%%%%%%%%%%%%%%%%%%%%%%
\section{Propagation of UHECRs in GMF}
%%%%%%%%%%%%%%%%%%%%%%%%%%%%%%%%%%%%%%%%%%%%%%%%%%
%%%%%%%%%%%%%%%%%%%%%%%%%%%%%%%%%%%%%%%%%%%%%%%%%%

In order to investigate the propagation of UHECRs in GMF, a backtracking method has been often adopted. In this method, particles with the charge opposite to that of particles which we are interested in are injected from the Earth and their trajectories are calculated. Then, we can regard the trajectories of these {\it anti-}particles as those of particles arriving from extragalactic space. In many cases, all the energy-loss processes of UHECRs are negligible because of the size of the Galaxy smaller than the energy-loss lengths of UHECRs. This method allows us to save much CPU time, since only the particles reaching the Earth are taken into account. This method can be applied to simulate the arrival distribution of UHECRs taking GMF and even structured IGMF into account \cite{Yoshiguchi:2003mc,Takami2006ApJ639p803}.

\begin{figure}
\includegraphics[clip,width=\linewidth]{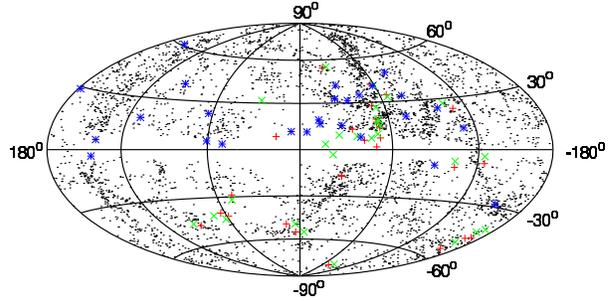}
\caption{Arrival directions of the 27 PAO events \cite{Abraham2007Aph29p188} ({\it red}) and their back-tracked directions assuming protons ({\it green}) and irons ({\it blue}) in galactic coordinates. A BS model with symmetry above and below the Galactic plane is assumed. Black points are galaxies in the IRAS catalog within 75 Mpc \cite{Saunders2000MNRAS317p55}. This figure is originally from Ref.\cite{Takami2010ApJ724p1456}.}
\label{fig:PAO_GMF}
\end{figure}

Since the solar system is embedded in the Galactic disk 8.5 kpc away from the Galactic center, magnetic field in the vicinity of the solar system mainly contributes to the total deflections of UHECRs in Galactic space. Fig. \ref{fig:trajectories} shows the examples of the trajectories of UHE protons ({\it left}) and irons ({\it right}) with the energy of $10^{19.8}$ eV with the same arrival directions ($\ell$, $b$) $=$ ($0.0^{\circ}$, $-44.8^{\circ}$). The GMF models adopted in this figure are BS and AS models for spiral field in the disk with two exponential decay scale in the direction perpendicular to the Galactic disk. The symbols S and A means symmetric and anti-symmetric field below and above the Galactic plane. The trajectories of protons seems almost to be straight lines by eye, but are deflected several degrees mainly in the Galactic disk. The field reversals of the disk field affect the total deflection of protons. In the BS models, the nearest field reversal is located at 0.5 kpc inside the solar system. When a charged particle passes through the field reversal, the direction of its deflection becomes opposite, and the total deflection angle is suppressed \cite{AlvarezMuniz:2001vf}. Therefore, BS models, in general, predict smaller deflection angles of UHECRs than AS models. We can observe this effect even for irons (green line), but sometimes heavy nuclei are confined in the Galactic disk even in the highest energy range. The propagation trajectory of a particle shown by the red line is very complex and is confined in the Galactic disk for a long time, and therefore directional information on its source is completely lost. The distribution of the deflection angles of UHE irons was well studied by Refs. \cite{MedinaTanco:1997rt,Giacinti2010JCAP08p036}, similarly to proton cases \cite{Takami:2007kq}, demonstrating that irons with the energy of $6 \times 10^{19}$ eV are deflected more than $25^{\circ}$ in more than 80\% region of the sky assuming a GMF model proposed by Ref. \cite{Prouza:2003yf} plus weak dipole field. Ref. \cite{Takami2010ApJ724p1456} calculated the back-tracked positions of UHECRs observed by the Pierre Auger Observatory (PAO) onto the surface of Milky Way sphere on the assumption of pure iron composition and demonstrated the correlation between UHECRs and nearby galaxies \cite{Abraham2007Sci318p938} is strongly disturbed. Fig. \ref{fig:PAO_GMF} is the arrival directions of 27 UHECRs above $5.7 \times 10^{19}$ eV in the first public data of the PAO \cite{Abraham2007Aph29p188} ({\it red}), the back-tracked directions of them on the assumption of protons ({\it green}) and irons ({\it blue}). A BS-S model is adopted. The black points are galaxies in the IRAS catalog within 75 Mpc \cite{Saunders2000MNRAS317p55}. While the back-tracked directions are close to the original arrival directions in the case of protons, the back-tracked directions of irons are far away from the original directions and no longer correlate with the distribution of galaxies. Pure iron composition at the Earth is an extreme and unrealistic case even in the case when all the UHECRs emitted from sources are irons because photodisintegration with cosmic microwave background during propagation in intergalactic space. Nevertheless, it is an intriguing problem whether the correlation can be reproduced under heavy-nuclei dominated composition if the correlation is not a statistical fake. 

As mentioned above, the spiral field in the vicinity of the solar system mainly contribute to the total deflections. Thus, reflecting the direction of local magnetic field, UHECRs from extragalactic sources in the northern Galactic hemisphere are deflected to the direction of Galactic south. This tendency can be seen in the figure. In the case of anti-symmetry above and below the Galactic plane, the deflection directions are opposite. In both cases, since the Larmor radius of a charged particle is proportional to its energy, the arrival directions of UHECRs from a source are arranged in the order of their energy if the coherent component of GMF is dominated \cite{AlvarezMuniz:2001vf,Yoshiguchi:2003mc,Takami2010ApJ724p1456}. 

The Liouville's theorem leads the isotropic distribution of UHECRs at the Earth if UHECR flux is isotropic outside GMF. However, the distribution of the momentum directions of back-tracked UHECRs injected from the Earth isotropically is not isotropic because there are the trajectories of UHECRs which cannot reach the Earth \cite{AlvarezMuniz:2001vf,Yoshiguchi:2004kd,Takami:2007kq,Giacinti2010JCAP08p036}. In other words, there are directions in which sources can provide the Earth with UHECRs efficiently (or inefficiently). Such an effect is called magnetic lensing (de-lensing) \cite{Harari2002JHEP03p045,Kachelriess2007APh26p378}. Interestingly, this effect can occur even in turbulent magnetic field in a few \% of the whole sky. 

%%%%%%%%%%%%%%%%%%%%%%%%%%%%%%%%%%%%%%%%%%%%%%%%%%
%%%%%%%%%%%%%%%%%%%%%%%%%%%%%%%%%%%%%%%%%%%%%%%%%%
\section{Searching for UHECR sources}
%%%%%%%%%%%%%%%%%%%%%%%%%%%%%%%%%%%%%%%%%%%%%%%%%%
%%%%%%%%%%%%%%%%%%%%%%%%%%%%%%%%%%%%%%%%%%%%%%%%%%

The correlation between UHECRs and the positions of source candidates has been expected to give a hint to identify UHECR sources, and therefore a lot of efforts has been dedicated. Although the modifications of the arrival directions by GMF are not taken into account to avoid model dependence in many studies, several tests considered plausible GMF models and obtained positive results. In early days, Ref. \cite{Tinyakov:2001ir} tested the correlation between observed UHECR events and BL Lac objects taking a BS model into account. The authors treated the charge of the UHECRs as a free parameter, and obtained a positive correlation signal in the case of $Q=+1$ contraty to no signal in the case of $Q=-1$, where $Q$ is the charge of UHECRs in the unit of the absolute value of the electron charge. This implies that the composition of UHECRs is mainly protons. Recently, Ref. \cite{Jiang2010ApJ719p459} examined the correlation between the PAO events and $\gamma$-ray sources detected by Fermi Large Area Telescope \cite{Abdo2010ApJ715p429}. Although strong ($\sim 4\sigma$) correlation was found, the modification by GMF did not largely change the result because a Cen A region is dominated in the signal. 

Ref. \cite{Nagar2010AA523A49} tested the positional correlation between observed UHECR events and galaxies with extended radio jets taking modifications of their arrival directions by several GMF models. The authors claimed that one third of UHECRs correlates with extended radio jets if they are light nuclei and proposed the idea that the remaining two thirds are heavy nuclei, i.e., a mixed composition scenario. If radio galaxies are responsible for the observed UHECRs, a BS-S or BS-A model is supported. Interestingly, several trajectories of UHECRs pass close to Galactic magnetars and microquasars under the BS-S model. The propagation of UHECRs from Galactic sources was studied by Ref. \cite{AlvarezMuniz:2005iy}. UHECRs with $10^{18}$-$10^{19}$ eV from Galactic sources inside the solar system arrive from the northern Galactic sky and from the southern Galactic sky for a AS-S and BS-S models, respectively. In the case that the A-type parity of GMF is realized, the difference between the two models is not clear. These results can easily scale to the case of heavy nuclei at the highest energy by using the rigidity scaling. Thus, large-scale anisotropy is expected at the highest energy if heavy nuclei are dominated and the S-type parity of GMF is realized.

\begin{figure}
\includegraphics[clip,width=\linewidth]{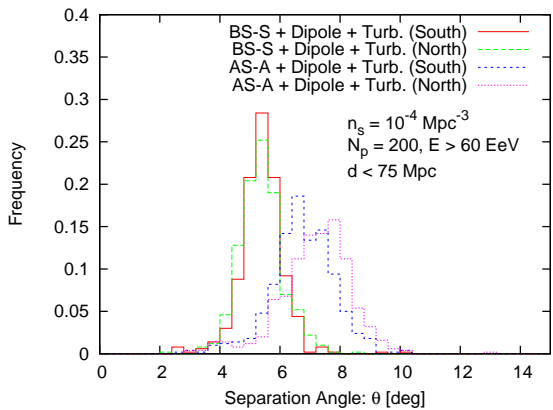}
\caption{Histograms of the separation angles between 200 simulated protons above $6 \times 10^{19}$ eV and their sources with $10^{-4}$ Mpc$^{-3}$ within 75 Mpc which gives maximal significance against random event distribution. The peak positions depend on the field reversals of GMF. This figure is originally from Ref.\cite{Takami2010ApJ724p1456}.}
\label{fig:thetabest}
\end{figure}

Another approach is to estimate typical angular scale within which positive correlation between UHECRs and their sources appears by simulations. This angular scale can be used for the cross-check of the results from correlation studies between observed UHECRs and the distribution of source candidates (galaxies), because we do not know what objects are UHECR sources. In addition, since the typical angular scale depends on composition, the estimation also gives an important information to constrain composition indirectly. Ref. \cite{Takami2010ApJ724p1456} simulated the arrival distribution of protons taking GMF into account and estimated typical angular scale within which positive correlation between the simulated protons and nearby sources used in the simulations. Fig. \ref{fig:thetabest} shows one of the histograms of the angular scale within which the correlation signal is maximized in the case of the number density of UHECR sources of $n_s = 10^{-4}$ Mpc$^{-3}$. Results a bit depend on GMF models and the number density of UHECR sources $n_s$. The typical angular scale is $\sim 5^{\circ}$ for $n_s = 10^{-4}$ Mpc$^{-3}$ and $\sim 6^{\circ}$ for $n_s = 10^{-5}$ Mpc$^{-3}$ in the cases of BS models. AS models predict angular scale larger than those in BS models due to larger deflections in AS models. The smaller angular scale is predicted in the larger number density because the cross-correlation function includes correlation between simulated UHECRs and sources not emitting the simulated UHECRs located in the vicinity of the sources contributing to the observed (simulated) events.

Motivated by ordered deflections by the regular component, the reconstruction of GMF structure along the line of sight and of the positions of sources has been studied by simulations \cite{Harari2002JHEP07p006,Golup2009APh32p269}. A recent study concluded that about ten events from a source above $3 \times 10^{19}$ eV allowed to reconstruct the source position with an accuracy of $0.5^{\circ}$ and the orthogonal component of the magnetic field with respect to the line of sight with an accuracy of 0.6 kpc Z$^{-1}$ \cite{Golup2009APh32p269}.

Candidates of the arranged events were found in the data recently published by the PAO in a Cen A region \cite{2010arXiv1011.6333G}. The source position reconstructed by the arranged events is located at $8.5^{\circ}$ from M87, a nearby powerful radio galaxy. According to a statistical test, the probability that such a set of events are realized in random background is $\sim 3 \times 10^{-5}$. This scenario offers the possibility that the excess of UHECR events around Cen A is not produced by Cen A, and gave us not only indirect evidence that M87 is a nearby source of UHECRs but also a possible solution to the fact that highest energy events do not seem to correlate with the position of the Virgo cluster \cite{Gorbunov:2007ja,Takami2008JCAP06p031}.

GMF produces the time-delay of UHECRs compared to neutral particles emitted at the same time. The time-delay plays an important role if the production of UHECRs is transient. For transient sources, this time-delay makes the apparent duration of UHECR bursts and we observe the sources as apparently steady sources in the observation time-scale of human beings \cite{MiraldaEscude1996ApJ462L59,Murase2008ApJ690L14}. Combined by the estimation of the number density of UHECR sources on the assumption of steady sources \cite{Takami2009Aph30p306,Cuoco2009ApJ702p825}, the time-delay by GMF leads an upper limit of the rate of UHECR bursts as (60-3000) Gpc$^{-3}$ yr$^{-1}$ \cite{Murase2008ApJ690L14}. Although the range means that this upper limit depends on the modelling of GMF, this limit rules out several transient phenomena for UHECR sources.

%%%%%%%%%%%%%%%%%%%%%%%%%%%%%%%%%%%%%%%%%%%%%%%%%%
%%%%%%%%%%%%%%%%%%%%%%%%%%%%%%%%%%%%%%%%%%%%%%%%%%
\section{Summary}
%%%%%%%%%%%%%%%%%%%%%%%%%%%%%%%%%%%%%%%%%%%%%%%%%%
%%%%%%%%%%%%%%%%%%%%%%%%%%%%%%%%%%%%%%%%%%%%%%%%%%

We have reviewed the effects of GMF to the arrival directions of UHECRs. The importance of GMF is in the fact that all the UHECRs arriving at the Earth are affected by GMF. A remarkable feature of GMF is its regular component. The regular component deflects the trajectories of UHECRs effectively and therefore significantly contributes to the arrival directions of UHECRs despite the smaller size of the Galaxy than the propagation distance of UHECRs in intergalactic space. The dominance of the regular component in UHECR deflections also allows us to use UHECRs as a spectrograph of GMF. Combined with increasing number of detected UHECR events and the improvement of the understanding of UHECR composition, the studies of UHECR propagation in GMF will help us understand the origin of UHECR sources. 

\begin{theacknowledgments}
H.T. thanks to the organizers of this workshop for inviting me. 
H.T. is partially supported by Grants-in-Aid for Scientific Research 
from the Ministry of Education, Cluture, Sports, Science and Technology 
of Japan through No.19104006 (through Katsuhiko Sato). 
\end{theacknowledgments}

\bibliographystyle{aipproc.bst} 
\bibliography{ms}

\end{document}